\documentclass[article,graybox]{svmult}

\usepackage{mathptmx}       % selects Times Roman as basic font
\usepackage{helvet}         % selects Helvetica as sans-serif font
\usepackage{courier}        % selects Courier as typewriter font
\usepackage{type1cm}        % activate if the above 3 fonts are
\usepackage{makeidx}         % allows index generation
\usepackage{graphicx}        % standard LaTeX graphics tool
\usepackage{subfigure}  
\usepackage{caption}                           
\usepackage{multicol}        % used for the two-column index
\usepackage[bottom]{footmisc}% places footnotes at page bottom
\usepackage{amsmath}

\makeindex             % used for the subject index
\long\def\omitit#1{}
\begin{document}

\title*{Size Agnostic Change Point Detection Framework for Evolving Networks}
\author{Hadar Miller and Osnat Mokryn}
\institute{Hadar Miller \at Information and Knowledge Management, University of Haifa, Israel\\ \email{name@email.address}
\and Osnat Mokryn \at Information and Knowledge Management, University of Haifa, Israel\\ \email{name@email.address}}
\maketitle
\omitit{
\abstract*{Each chapter should be preceded by an abstract (10--15 lines long) that summarizes the content. The abstract will appear \textit{online} at \url{www.SpringerLink.com} and be available with unrestricted access. This allows unregistered users to read the abstract as a teaser for the complete chapter. As a general rule the abstracts will not appear in the printed version of your book unless it is the style of your particular book or that of the series to which your book belongs.
Please use the 'starred' version of the new Springer \texttt{abstract} command for typesetting the text of the online abstracts (cf. source file of this chapter template \texttt{abstract}) and include them with the source files of your manuscript. Use the plain \texttt{abstract} command if the abstract is also to appear in the printed version of the book.}}

\abstract{Changes in the structure of observed social and complex networks' structure can indicate a significant underlying change in an organization, or reflect the response of the network to an external event. Automatic detection of change points in evolving networks is rudimentary to the research and the understanding of the effect of  such events on networks. Here we present an easy-to-implement and fast framework for change point detection in temporal evolving networks. Unlike previous approaches, our method is size agnostic, and does not require either prior knowledge about the network's size and structure, nor does it require  obtaining historical information or nodal identities over time. We use both synthetic   data derived from dynamic models and two real datasets: Enron email exchange and Ask-Ubuntu forum. Our framework succeeds with both precision and recall and outperforms previous solutions.}

\section{Introduction}
\vspace{-.4cm}
Complex systems of interacting elements, from human (social and organizational) to physical and biological ones, can be modeled as interaction networks, with nodes representing the elements and edges representing their interactions. When the interactions are dynamic, i.e., human and social interactions, a complete model that captures the longitudinal evolution of the system is comprised of a sequence of networks, each portraying a snapshot of the system at a single point in time.

Of specific interest recently is the analysis of changes in dynamic social and complex networks in response to events, and the automatic detection of these points of change, termed Change Point Detection (CPD). Recent works identified changes in the community partitioning of the Enron email exchange immediately after the Californian blackouts~\cite{Peel2015}, and a turtling up of conversation networks between traders in response to significant stock price changes~\cite{romero2016social}. Understanding the network's reaction to unusual events provides improved abilities to analyze, understand and possibly take actions in a given system, infer its reaction to external shocks, and aid in predicting organizational and behavioral changes.

Past research for identifying change points used stochastic models, of either scalar values representing the longitudinal data~\cite{McCulloh2011}, or probabilistic and model-based representations of the network~\cite{koutra2013deltacon, Peel2015, wang2017fast}, and did not examine the complex network's structure as manifested through distributions.

The structural properties that are in the focus of our work here are the network's native statistical distribution, i.e.,  its degree distribution measure. Distribution functions are a measure of the division of resources within the network, and their relative positions, and are considered a fundamental tool in the understanding of complex systems. Stumpf and Porter~\cite{Stumpf2012} have discussed this notable role,  claiming that %statistically sound fits of empirical data to a theoretical 
degree distributions %has practical implication as an 
serve as an aiding tool for understanding, interpreting and even predicting behaviors in a given system. Bhamidi et al.~\cite{bhamidi2018change} further showed that degree-distribution measures reflect changes in the underlying structure better than  the hyper-parameters of the corresponding network models. 

An additional valuable advantage of a degree distribution-based event detection is that it eliminates the need to know in advance the number of nodes in the network at each point in time, and can work with as little information as the sequence of interactions for the periods under inspection.   Thus, unlike all previous CPD schemes, the proposed solution here assumes no prior-knowledge of the network, does not require pre-processing, and can be used in an online manner, where new network snapshots are generated on-the-fly.

Here, we devise an online fast change point detection mechanism, utilizing the degree distribution of snapshots of networks in time. The detection mechanism is planned in a manner that does not require to determine exact theoretical fits to the distributions. We conduct a hypothesis testing to assess the significance of the change and differentiate a change signal from local fluctuations.  %facilitating {\em fast} detection and , we do not compute a fit for the degree distributions. Rather, we estimate the probability for a change between two consecutive network snapshots, conducting a hypothesis testing. %This is calculated as the  probability that the distribution measured in the latter snapshot could have been drawn from previous snapshot's distribution function,  using a hypothesis testing with a bootstrapping Kolmogorov-Smirnov distance measure.
%The hypothesis testing approach enables us to address the  difficult problem of identifying statistically significant change that is a result of a shift in the norm, and differentiate it from a mere noise~\cite{Peel2015}. 

The contributions of the work are the following:
\vspace{-.08in}
\begin{enumerate}
\item Taking a sliding-window approach for the network interactions, this method can address both the anomaly detection problem, in which there is a significant variation from a norm, and the change point detection problem, which considers a significant change to the norm itself, by computing the significance measure of the change (calculated p-value) over different window sizes.  
\item The approach is the first that does not require to know in advance the number of interacting nodes in each stage of the network's life, and hence can be used online.
\item We investigate the performance of the scheme over both synthetic data and real world data. For the synthetic data we conduct a thorough investigation  of several generative models, i.e.,  random networks and small world networks, with varying rate of events and over different network sizes.  This enables us to quantify the reaction of different network models to events. We further show that over two real datasets, the scheme performs better than existing detection schemes, while being faster. %Real datasets inspected are the Enron email, and the forum of conversation discussing Ubuntu Linux.   
\item The hypothesis testing we conduct enables a sensitivity measure for a change. First, it enables to set the level of sensitivity of a change according to need. Then, it opens the possibility to detect changes with decreasing sensitivity during a window of time. While current schemes detect reactions to shocks, this scheme can detect gradual changes that follow a clear trend of increasing probability of a change and can be utilized as a predictive framework.  	
\end{enumerate}
 
In section~\ref{sec:bg} we give the background and discuss  change point detection mechanisms. We then describe our detection framework in Section~\ref{sec:fw}. In Section~\ref{sec:res1} we review the performance of the framework over different generative models using synthetic data, and describe our results over two real datasets. We discuss our conclusions in Section~\ref{sec:con}.

\section{Background and Related Works}
\label{sec:bg}
\vspace{-.4cm}
It is widely accepted that structural properties of a network play a significant role in determining its actors' behavior~\cite{granovetter1983strength, burt2000network, haynie2001delinquent, spencer2003global,kossinets2006empirical, fowler2008dynamic}. The last decade's abundance of temporal information paved the path to a further understanding of the dynamics of networks~\cite{lazer2009life}, and  findings corroborate that structural properties have a prominent affect on the longitudinal dynamics of  networks and their actors~\cite{kossinets2006empirical,Leskovec2007,fowler2008dynamic, phelps2010longitudinal,ilany2015topological}.

In this work we investigate the effect of events on social networks. Romero, Uzzy, and Kleinberg in a recent novel work~\cite{romero2016social} defined these events as mostly exogenous events  that are either unexpected, or are extreme, relative to the average~\cite{romero2016social, gilbert2005unbundling}. %and investigated the effect of sudden change in a stock price on the social network of traders. 
%They created a daily temporal snapshot of the social network of traders discussing each stock (a network per stock per day), and examined some of the structural and behavioral changes in these networks after extreme external events (shocks). 
They found a turtling-up of the network as a reaction to an external shock, and measured changes  in the clustering coefficient, tie strength and percentage of border edges.  Kondor et al.~\cite{1367-2630-16-12-125003} researched the longitudinal structure of the  network of the most active Bitcoin users for a period of two years, and searched for important changes in the graph structure by comparing successive snapshots of the active core of the transaction network using principle component analysis (PCA). They found a clear correspondence with the market price of Bitcoin. McCulloh \& Carley ~\cite{McCulloh2011} included in their analysis of change points also cases of endogenous changes, and showed that their detection system can determine that a change has occurred from a longitudinal analysis of the network itself. Using their method, Tambayong~\cite{tambayong2014change} examined Sudan's political networks and found that foreign-brokered signings of multiple peace agreements served as a political solidification point for political actors of Sudan during the recent violent domestic conflict. According to their analysis, this was a catalyst that caused three leaders to have emerged and lead the more compartmentalized yet faction-cohesive political networks of Sudan. In a recent analysis, Peel and Clauset~\cite{Peel2015} were able to detect  external changes during the Enron crisis through a stochastic analysis of the Enron organizational email exchange \cite{Klimt2004}.

 Considering that distributions in complex systems have practical importance as an aiding tool for data interpretation and event prediction~\cite{Stumpf2012, bhamidi2018change},  we investigate here the interplay between points of change and this fundamental structural distribution in social organizations and systems. %Bhamidi et al.~\cite{bhamidi2015change} have investigated the effect of events that cause structural changes by incorporating change points into a Preferential Attachment model, and found that change points affect the model's degree distribution.  
%\vspace{-.8cm}
\subsection{Models for Change Point Detection in Networks}
\vspace{-.4cm}
In stochastic  models of networks, change points are points in time where a change in the system's norm is detected in a manner that can be significantly differentiated from plain stochastic noise~\cite{Peel2015,akoglu2010event,hirose2009network,gupta2014outlier,akoglu2015graph}.
 McCulloh \& Carley ~\cite{McCulloh2011} convert the series of networks to a time series of scalar values for different network measures, and looked for a  stable change in these values (as opposed to temporal change, when looking for anomaly detection) using process mining techniques for  change points detection~\cite{hawkins2003changepoint,priebe2005scan,McCulloh2011}. 
 
\omitit{Change point detection frameworks~\cite{akoglu2015graph} commonly consist of three  phases: (1) generate longitudinal graphs; (2) capture representative features of each graph; (3) detect significant distance  between  graph snapshots, or window of snapshots, to specify changes. } 

 Methods for CPD  differ mainly by the graph features they compute. A model-based approach fits each snapshot to a generative model. For example,  General Hierarchical Random graph  (GHRG), Generalized Two Block Erdos-Renyi (GBTER), and Kronecker Product Graph Model (KPGM)~\cite{Peel2015, bridges2015multi, moreno2013network}. A model-based approach requires a  pre-processing phase, for which  enough history is pre-known. It further requires that labeled nodal information is known. When taking the degree distribution, we eliminate the need for this extended information, as degree distribution does not require historical information, nor the node names. Moreover, recent analysis found that structural changes are better detected by the degree distribution than by the hyper-parameters of the generative model, for the PA case~\cite{bhamidi2018change}.
 
 A complementary approach, similar in nature to ours, is to extract a large number of features from each consecutive graph snapshots, and find the distance between them~\cite{akoglu2010event, koutra2013deltacon, wang2017fast, donnat2018tracking}. A change is determined if a predefined threshold for the distance is crossed. 
 
Unlike previous works that consider graph features, in our work, we conduct a hypothesis test to determine a change, to provide a certainty level for a change point detection.

\section{Detecting Change Points in Networks}
\label{sec:fw}
\vspace{-.4cm}
We explain our method following Figure~\ref{fm1}. A sequence of networks is presented, where a change in the generative model occurs. The change is not tied to a specific structural characteristic. Our framework computes the cumulative distribution function of the degrees (CDF) for each graph, computes the distance, and performs a hypothesis testing to infer how probable is a change given the measured distance between the two CDF's. Here, we chose the nonparametric Kolmogorov-Smirnov (KS) two-sample test to measure the distance, though other non-parametric statistical methods for measuring the distance may be applied.

\begin{figure}[t!]
%\centering{
\includegraphics[scale=.3]{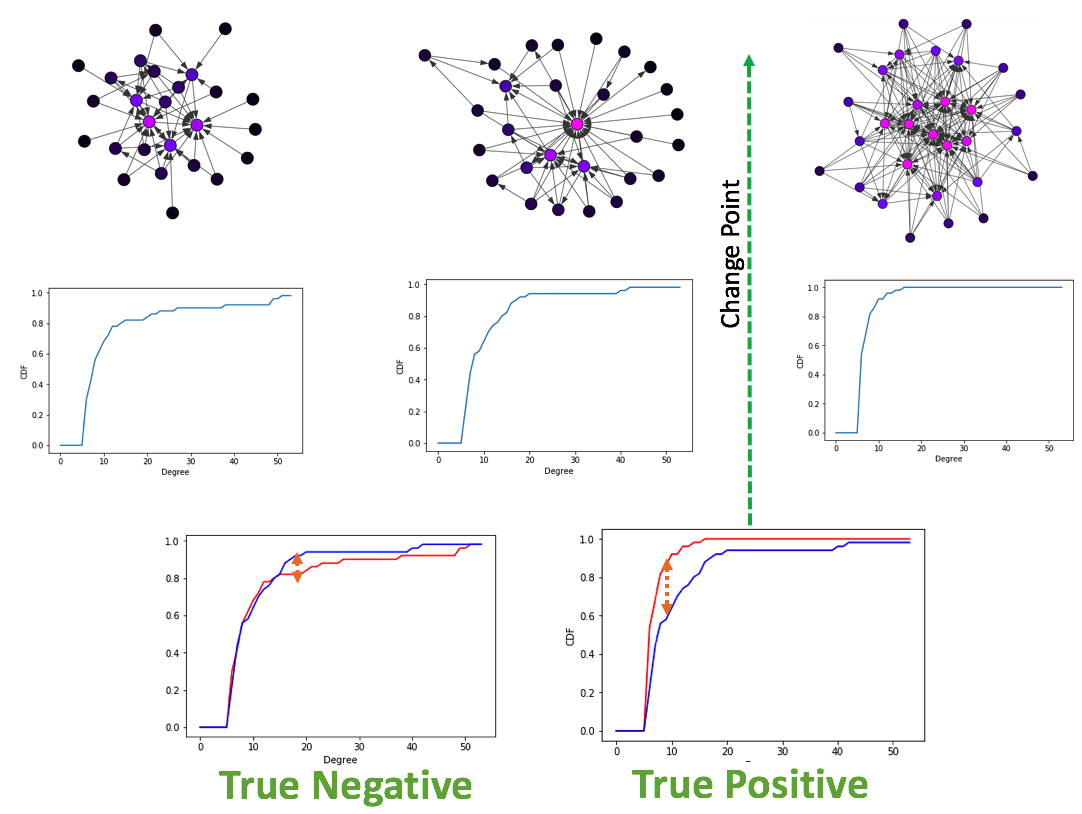}
\caption{\small{Our framework for detection of changes: Window size defines the stability of the change over time. Hypothesis testing over a distance measure is used to determine whether the underlying model has changed. On the left graphs generated from the same model, on the right a graph generated from a changed model}}
%}
\label{fm1}
\vspace{-.4cm}       % Give a unique label
\end{figure}

\begin{figure}
\centering{	
%\end{figure}
\begin{minipage}{.49\textwidth}
 %\centering{
  \includegraphics[width=\linewidth]{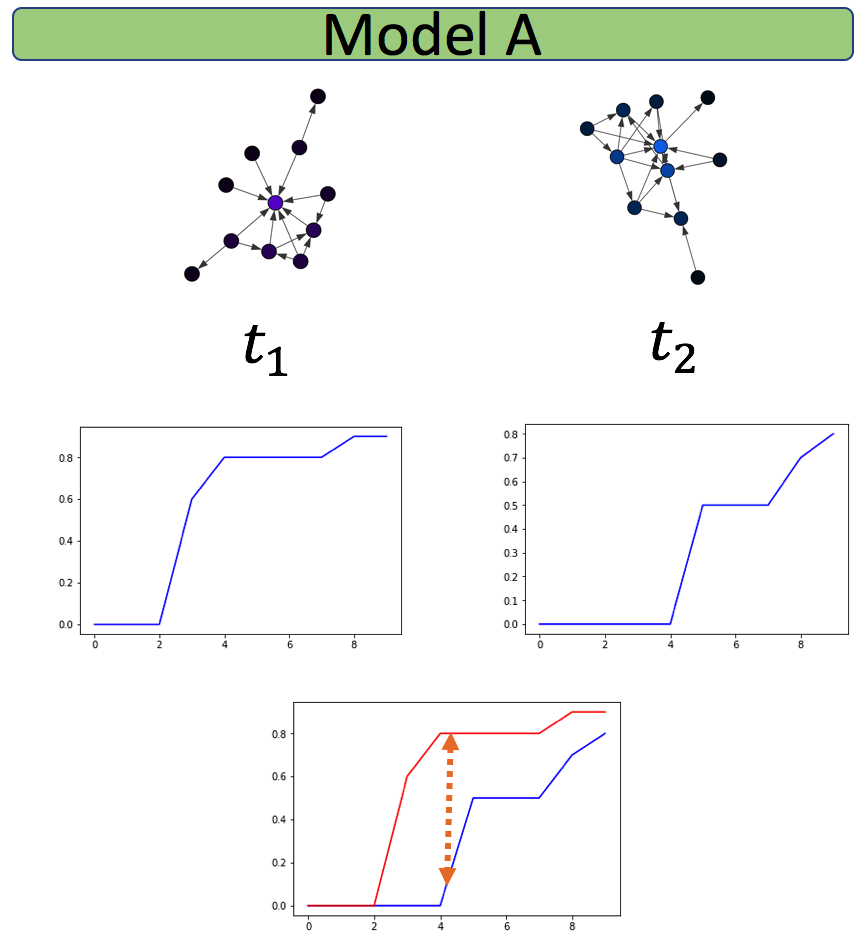}
      \vspace{-2mm}
    \captionsetup{justification=centering,margin=.1cm}
  \caption{False Positive: Distance measure is large as sample size is too small, although graphs come from same generative model}
%  }
\label{fig:fp}
\end{minipage}%
%\begin{figure}[!h]
\hspace{1mm}
\begin{minipage}{.49\textwidth}
 % \centering{
  \includegraphics[width=.97\linewidth]{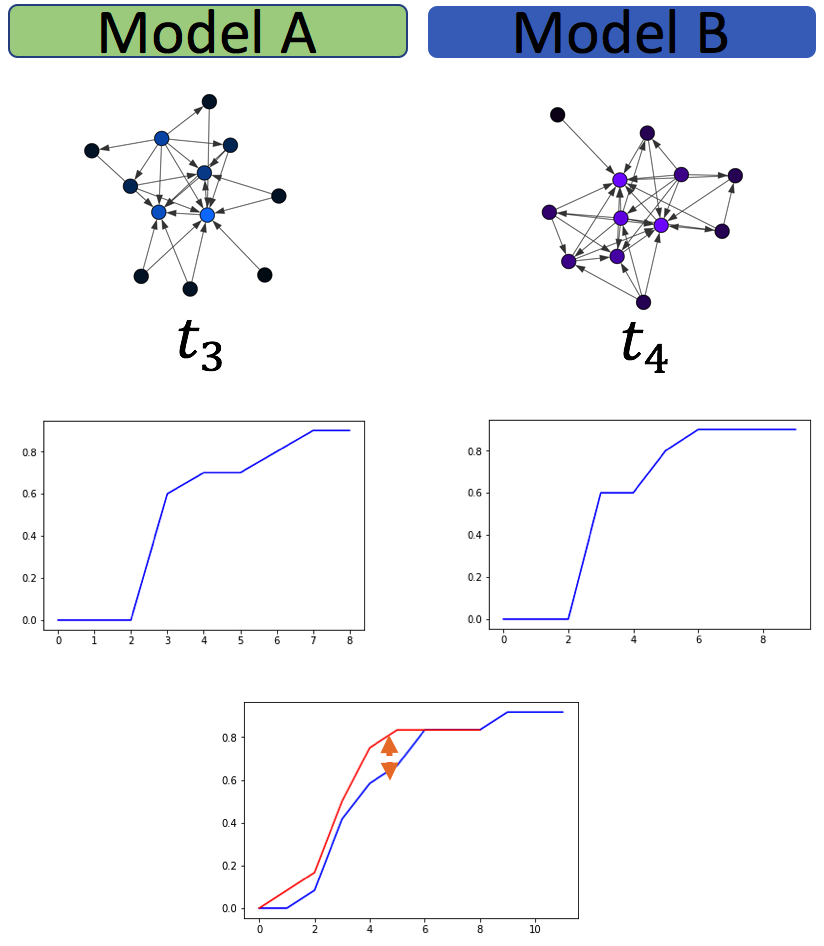}
     \vspace{-2mm}
  \captionsetup{justification=centering,margin=.1cm}
  \caption{False Negative: Fluctuations conceal each other and decrease the measured distance between  two networks}
%  }
  \label{fig:fn}
\end{minipage}
}
%\vspace{-.4cm}
\end{figure}

\begin{figure}
\vspace{.5cm}
%\centering{
%\begin{minipage}{\textwidth}
% \centering{
  \includegraphics[width=.98\linewidth]{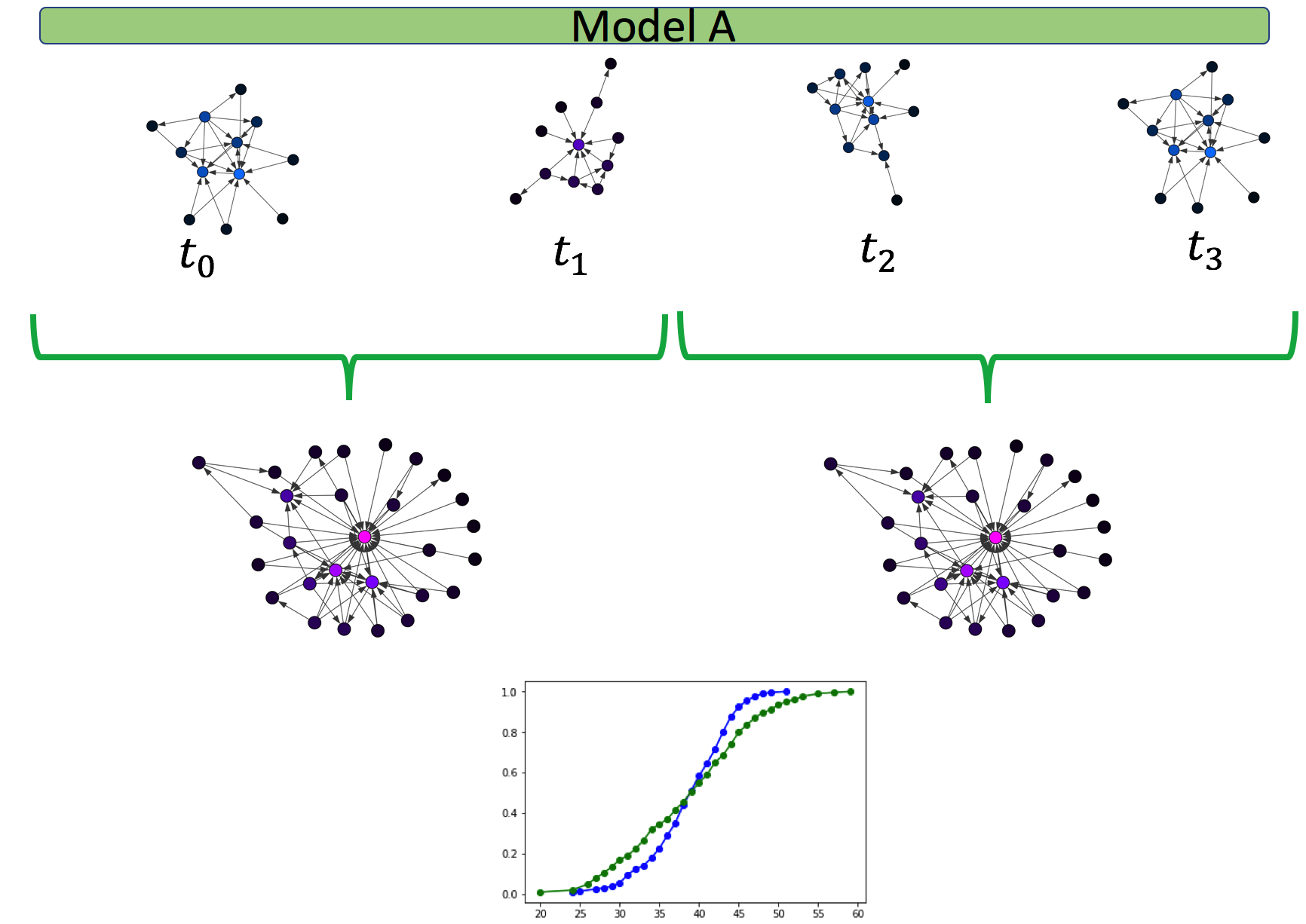}
      \vspace{-2mm}
  \caption{The use of a sliding window over several graph snapshots decreases the probability of a false positive estimation of a change}
%  }
\label{fig:sw}
% \end{minipage} }
\vspace{-.4cm}
\end{figure}

CPD frameworks as the ones discussed in the previous section divide the data to consecutive snapshots according to a natural devision derived from the nature of the data, such as daily or weekly snapshots of organizational frameworks, or monthly graphs of votes. In methods measuring the distance between features extracted from two consecutive graph snapshots~\cite{akoglu2010event, koutra2013deltacon, wang2017fast}, a change is detected if the measured distance is bigger than an arbitrarily predefined threshold value. But distance measures work well mainly for large sample sizes. When the sample size is small, a large distance can be measured, crossing the threshold value. This can lead to a false positive result that a change occurred, when there is merely a fluctuation in the network that should be identified as  noise, and is considered a false positive inference, as demonstrated in Figure~\ref{fig:fp}. To avoid these types of false inference we suggest the use of a sliding window over several graph snapshots, and computing the CDF across the entire window, as is the case in Figure~\ref{fig:sw}. A complementary situation occurs when windows that are set too large conceal an event within them, thus hiding the point of change. This would correspond with a false negative inference, and is demonstrated in Figure~\ref{fig:fn}. A solution for this problem is the use of a sliding window to find the exact point of change within the window, as is used in~\cite{Peel2015}.  
An alternative approach to measuring a distance between windows is to try and fit a theoretical statistical distribution to each network snapshot, and  determine whether they are derived from the same model. This is, however,  a rather  time and computational-intensive approach. To fit data to a statistical theoretical model requires both to find a fit and to reject other possible theoretical statistical distributions~\cite{Clauset2009}. Hence, we compare distances across windows, as described in Figure~\ref{fig:sw}.

We conduct a hypothesis testing for understanding whether the distance between the degree distributions asserts that they come from the same model, or from two different generative models.  %the distance measure between the windows, as explained here.
We measure the distance between the cumulative degree distributions of consecutive snapshots. For any two  consecutive windows, let us define their graphs representations as $g_i,g_{i+1}$. The null hypothesis is that the cumulative distributions measured for any two consecutive snapshots, $g_i, g_{i+1}$, are drawn from the same distribution, $G_{Null}$, in which case no change has occurred between the windows. To test the hypothesis we generate synthetic datasets from the distribution of $g_i$ and find their distributions. The standard approach for generating samples for hypothesis testing is bootstrapping, which generates samples by randomly re-sampling (with replacement) the data ~\cite{efron1994introduction}.  We use here the nonparametric Kolmogorov-Smirnov (KS) two-sample test. The method is considered robust and is widely used. Yet, when comparing two distributions using too few samples it can fail to reject a false null hypothesis.  The probability for that diminishes as the number of nodes interacting in each snapshot increases, and it is best to create snapshots that contain, as a rule of thumb, at least 50 nodes each.

%to In the heart of our framework is a hypothesis thesis for determining whether a change occurs, causing a change in the network structure between two consecutive snapshots. A change is detected if currently inspected network dynamics, as captured by the degree distribution, deviate  significantly from the previous ones. Clearly, interactions change in time, and local changes occur often. These are normal dynamics of complex networks, and should not be inferred as a change. Rather, we look for a significance change, signaling that an event has altered significantly the dynamics of the interactions.  
 
%We repeatedly measure the distance between the cumulative distributions of using the nonparametric Kolmogorov-Smirnov (KS) two-sample test. 
 
For two consecutive graph snapshots \(g_i, g_{i+1}, (i \in \{1,2,...\})\) we denote the two generated corresponding cumulative degree distribution functions by \(S_i(x), S_{i+1}(x)\). 
Given the CDF degree distribution $S_j(x), j \in {i, i+1}$ for graph $g_j$:
\(S_j(x)=P_j(x \leq X)\) we compute the KS statistic $D$, defined as the maximal difference between the two empirical distributions, as described by Equation~\ref{eqn:1}. 
The KS null hypothesis is that the two samples where drawn from the same distribution. 
 \begin{equation}
\label{eqn:1}
D_{i,i+1} = \underset{x}{\sup} | S_i(x) - S_{i+1}(x)|
% D_i = \sup | |
\vspace{-.2cm}
\end{equation}
The KS null hypothesis is rejected with significant level $\alpha$ if the computed distance ($D_{i,i+1}$) is greater than some critical value.

As explained before, a large KS distance $D_{i,i+1}$ measured between $S_i(x)$ and $S_{i+1}(x)$ doesn't necessarily indicate a signal in our framework. We would like to test how rare such distance $D_{i,i+1}$ is. We define $g_i$ as the base model graph, and   
conduct a hypothesis testing, with a null hypothesis that the distance $D_{i,i+1}$ between the base model graph distribution and the consecutive one is not rare for samples taken from the same statistical model. 

Our null hypothesis then assumes that the distance between the snapshots' distributions is typical for distances between distributions sampled from the base model graph distribution. The null hypothesis is rejected with significance $p$ if in $(1-p)$ percent of the times the measured distance between $S_i(x)$ and the sampled distributions is smaller than $D_{i,i+1}$, as depicted in Equation~\ref{eqn:label}.

%We   %distance or bigger ones occur between  samples drawn from the distribution of $g_i$ compared distributions.  
Following the bootstrap procedure~\cite{efron1994introduction} we generate  $j=1000 >> 1$ new samples from $S_i(x)$ and measure the distance $d_{i,j}$ between $S_i(x)$ and each of its bootstrap samples. We test the hypothesis by  computing the fraction of times a KS test will yield a distance $D_{i,j}, j \in \{1..10000\}$, that is at least as big as $D_{i,i+1}$. \begin{equation}
\label{eqn:label}
{p} = \frac {|D_{i,i+1} > \{d_{i,j}\}|}{|\{d_{i,j}\}|} 
\end{equation}
A confidence level $\alpha$ may now be chosen to reject the null hypothesis, depending on the acceptable false positive rate.
This confidence level corresponds to the sensitivity of the change, and can be tunable.

% !TEX root = MMcpd.tex
\section{Detecting Changes over Different Network Types}
\label{sec:res1}
\vspace{-.4cm}
We conduct several experiments to evaluate the performance of our framework, on both large synthetic datasets and real networks. First, we investigate the performance of the framework on synthetic networks generated by several generative models. Each such generative model enables us to investigate the framework's behavior for different structural characteristics.  As our method is based on the degree distribution of the network it is agnostic to any changes in the network size. Hence,  we expect our framework to detect changes across network snapshots that may gain or lose nodes during the network's lifetime. At first we considered to use a preferential-attachment growing network as one of the models.  However, this model is specifically designed to explain the emergence of hubs in networks and  the long tail distribution of real-world networks degrees, and thus is designed to create a specific degree distribution, which is what we try to find. % effects the degree distribution by definition and though we preferred to test our framework using less trivial model. 
For generative models we therefore employ the Erd\"{o}s-R\'enyi (ER) random networks model and the Caveman model. %The latter enables us to explore in detail the framework over a community-structured network.
%;  and a growing preferential attachment network,  as detailed in Table~\ref{tbl:1}. All generated networks are dynamic. 
\omitit{
\begin{table}
\caption{Synthetic Networks: Generative Models and Structural Characteristics}
\label{tbl:1}       % Give a unique label
%
% Follow this input for your own table layout
%
\begin{tabular}{p{4cm}p{4.9cm}p{2.4cm}}
\hline\noalign{\smallskip}
Model & Structural Characteristics &  Degree Distribution  \\
\noalign{\smallskip}\svhline\noalign{\smallskip}
%Translation & mRNA$^a$  & 22 (19--25) & Translation repression, mRNA cleavage\\
Erdos-Renyi~\cite{} & Giant Component, Phase transition & Poisson\\
Caveman~\cite{}  & Small World, Connected Components & Normal \\
Preferential Attachment~\cite{} & Hubs, Scale Free & power law\\
\noalign{\smallskip}\hline\noalign{\smallskip}
\end{tabular}
%$^a$ Table foot note (with superscript)
\end{table}
}
In each experiment  the network model alternates between two configurations that differ in their hyper parameters. The number of changes is set to 100, distributed randomly. Then, the number of consecutive snapshots of the network drawn from the model configuration, $x$, is chosen from a normal distribution $x \sim N(\mu=4,\sigma^2=2)$, such that the average number of consecutive graphs in each experiment is on average 400.  \\[14pt]
%Every model type was investigated under a variety of conditions, explained for each model and also detailed in summary Table~\ref{tbl:results}. For each model configuration a sequence of dynamic networks is created in time, Each iteration of each model configuration contains 100 points of change distributed randomly in time.  interval $x ~^{iid}(N,4,2)$over the network timeline  
{\large Random Graphs:\\} We start with the Erd\"{o}s-R\'enyi (ER) random graph  model~\cite{ER60}. The model for random graphs $G(n,p)$ assumes a fixed number of nodes $n$\footnote{Recently Zhang et al.~\cite{zhang2017random} suggested a generalization for dynamic random networks, in which the dynamic process is governed by a continuous Markov-process. As we need to study the change point detection problem, requiring the change in the generative model hyper parameters, we could not employ their model.}. Edges connect node pairs independently with probability $p$. Low values of $p$ entail that the number of edges is substantially lower than the number of nodes, and the model generates small components in tree forms. 
As $p$ increases, and reaches  $p > o(\frac{1}{n})$, the network changes to suddenly form a giant component, a phase transition that has a distinct influence on the structure of the network. 

\begin{figure}[t]
%\samenumber
\subfigures
\begin{minipage}{.49\textwidth}
  \includegraphics[width=\linewidth]{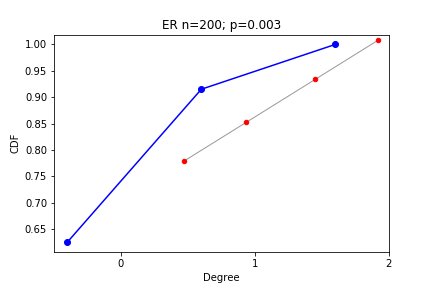}
    \vspace{-2mm}
    \captionsetup{justification=centering,margin=.4cm}
\caption{Random Network exp.1 (200 nodes) $1^{st}$ configuration ($p=.003$, fragmented) degree CDF and largest component}
   \label{fig:er11}
  \end{minipage}%
\begin{minipage}{.49\textwidth}
  \includegraphics[width=\linewidth]{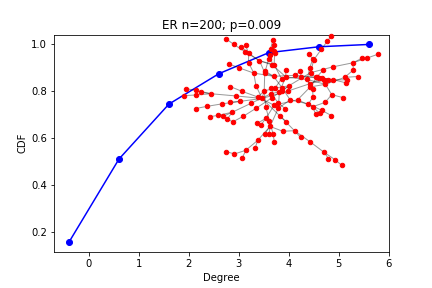}
  \vspace{-2mm}
      \captionsetup{justification=centering,margin=.4cm}
\caption{Random Network exp.1 (200 nodes) $2^{nd}$ configuration ($p=.009$, connected) degree CDF and largest component}
  \label{fig:er12}
\end{minipage}
%\\[3pt]
\begin{minipage}{.49\textwidth}
  \includegraphics[width=\linewidth]{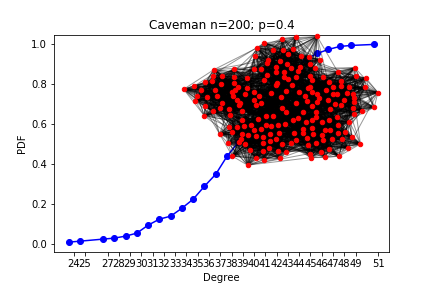}
    \vspace{-2mm}
    \captionsetup{justification=centering,margin=.2cm}
\caption{Caveman exp.3 (200 nodes, C=5) $1^{st}$ configuration ($p=0.4$) CDF and visualization}
   \label{fig:cave11}
  \end{minipage}%
\begin{minipage}{.49\textwidth}
  \includegraphics[width=\linewidth]{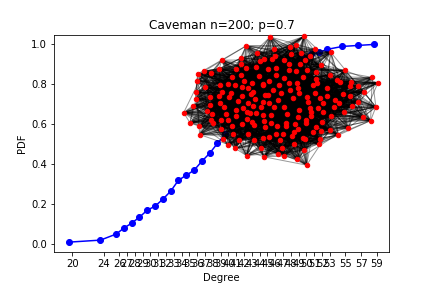}
    \vspace{-2mm}
      \captionsetup{justification=centering,margin=.1cm}
\caption{Caveman exp.3 (200 nodes, C=5) $2^{nd}$ configuration ($p=0.7$) CDF and visualization}
  \label{fig:cave12}
\end{minipage}
\label{fig:test}
\vspace{-.4cm}
\end{figure}

We then perform two experiments for this model type, as described here, and detailed in Figures~\ref{fig:er11},~\ref{fig:er12}:
%\vspace{4mm}

\begin{itemize}
\vspace{-2mm}
 \item Experiment 1 -  A change in the hyper-parameters of the ER model transitions the network between the two network states of fragmented ($p << o(\frac{1}{n})$) and connected ($p > o(\frac{1}{n})$). The networks configurations are the following. Each configuration consists of 200 nodes, and the model's hyper parameter is either $p=0.003$, i.e., fragmented, or $p=0.01$, i.e., connected.  
 \item Experiment 2 -  The ER networks consist of 200 nodes each, and the model's hyper parameter is either $p=0.1$, $p=0.15$, i.e., both times the network is connected, and there is a slight change in the connectedness. It is safe to assume that the subtleness of the change in the generative model of the random network will make it harder to identify the change.
 \end{itemize}
 
{\large Caveman Model: \\}
The ER model generates graph with small clustering coefficients, which lack the capability to represent communities. Social networks are often characterized as having highly connected communities that form rare interactions in between, and form a {\em Small World}. For example, in an organization you may expect intensive interactions between actors within departments and sparse interactions between actors belonging to different departments. A generative model for a small world network is the Caveman~\cite{watts1999networks}.

To test our framework against networks with varying sizes we generated a sequence of unlabeled networks, $g_i \in   G$, while using the Caveman model. The number of nodes for each snapshot was randomly selected from a uniform distribution $||g_i|| \sim U(200, 1000)$. To prevent a sample size bias while calculating the KS distance we randomly sampled 200 nodes from each network and calculated the distance between the two samples degree distribution.
\begin{itemize}
\vspace{-2mm}
\item Experiment 3 - The Caveman-based networks are drawn from the a model containing 200 nodes, as explained above, and $C=5$ communities each. The change in the hyper-paramter between the two configurations is in the rewire probability $p$. In the $1^{st}$ configuration, visualized in Figure~\ref{fig:cave11} $p=0.4$. In the $2^{nd} p=0.7$, leading to a more inter-connected network, as is visualized in Figure~\ref{fig:cave12}.
\end{itemize}

\vspace{-.7cm}
\paragraph{\large{Detection Performance}}
  \vspace{-.3cm}
%In ER networks the degree of a vertex follows a binomial distribution with success probability $p= \frac{1}{\lambda}$ and $n-1$ trails. Thus, 
Table~\ref{tbl:res} describes the performance of our detection framework for the three described experiments. Note, that for the ER networks (exp1, exp2) we get a perfect recall. The degree distribution of an ER random graph with edge probability $p= \frac{\lambda}{n}$ follows a Poisson distribution with probability mass function: $e^{-\lambda}\frac{\lambda^k}{k!}$, with mean $\lambda$ and  skewness $\lambda^{-0.5}$.  A change in  $\lambda$  differentiates two ER generative models and will be projected to the networks' CDF, thus detectable by our model.  This may explain the perfect detection (Recall=1) of all events in our synthetic data tests. However, the variance of a Poison distribution is $\lambda$ as well. As the variance $\lambda$ increases, the chances of mistakenly find two samples drawn from the same model as not sharing the same distribution increase. This explains our relative low precision. %More specifically in our tests the precision decreased as we increased in the edge probability between the tests. Theoretically we could increase our precision by reducing the $\lambda$ as low as possible. However, trying to reach values lower that 1 puts the network generative model far from its transition phase. 

As true positive events (change points) were detected with significance that exceeds $99\%$ We repeated the experiments while increasing the CPD threshold from $90\%$ to $99\%$. This test resulted with Recall = 1.0 and Precision = $0.89$. This corresponds to changing the {\em sensitivity} of framework, as discussed before.
\begin{table}[h]
\vspace{-.6cm}
\caption{CPD Framework Performance For Synthetic Networks}
\label{tbl:1}       % Give a unique label
%
% Follow this input for your own table layout
%
\vspace{-.2cm}
\begin{tabular}{p{1.8cm}p{6cm}p{2.5cm}p{2.5cm}}
\hline\noalign{\smallskip}
Experiment &Model \& Main Structural Property &  Precision & Recall  \\
&& Mean, Std & Mean, Std \\
\noalign{\smallskip}\svhline\noalign{\smallskip}
%Translation & mRNA$^a$  & 22 (19--25) & Translation repression, mRNA cleavage\\
exp1 & ER: Phase transition $p=\{0.003,0.009\}$ & 0.767, 0.03&1.0, 0.0\\
exp2 & ER: Connected $p=\{0.1,0.15\}$  & 0.671, 0.02&1.0,   0.0\\
exp3 & Caveman: Communities $p=\{0.4,0.7\}$  & 1.0, 0 & 0.961, 0.01\\
\noalign{\smallskip}\hline\noalign{\smallskip}
\end{tabular}
\label{tbl:res}
%$^a$ Table foot note (with superscript)
\vspace{-.4cm}
\end{table}
 \vspace{-.4cm}

The results for the Caveman model (exp 3) yield excellent results of perfect precision ($100\%$) and near-perfect recall ($96\%$), showing that a community structure of networks lends itself naturally to our detection framework. 
 
%\vspace{-.8cm}
\subsection{Detection of Events Changing Real Networks}
\vspace{-.4cm}
 
\begin{figure}[t]
%\samenumber
%\subfigures
\begin{minipage}{\textwidth}
\hspace{-1cm}
  \includegraphics[width=1.2\linewidth]{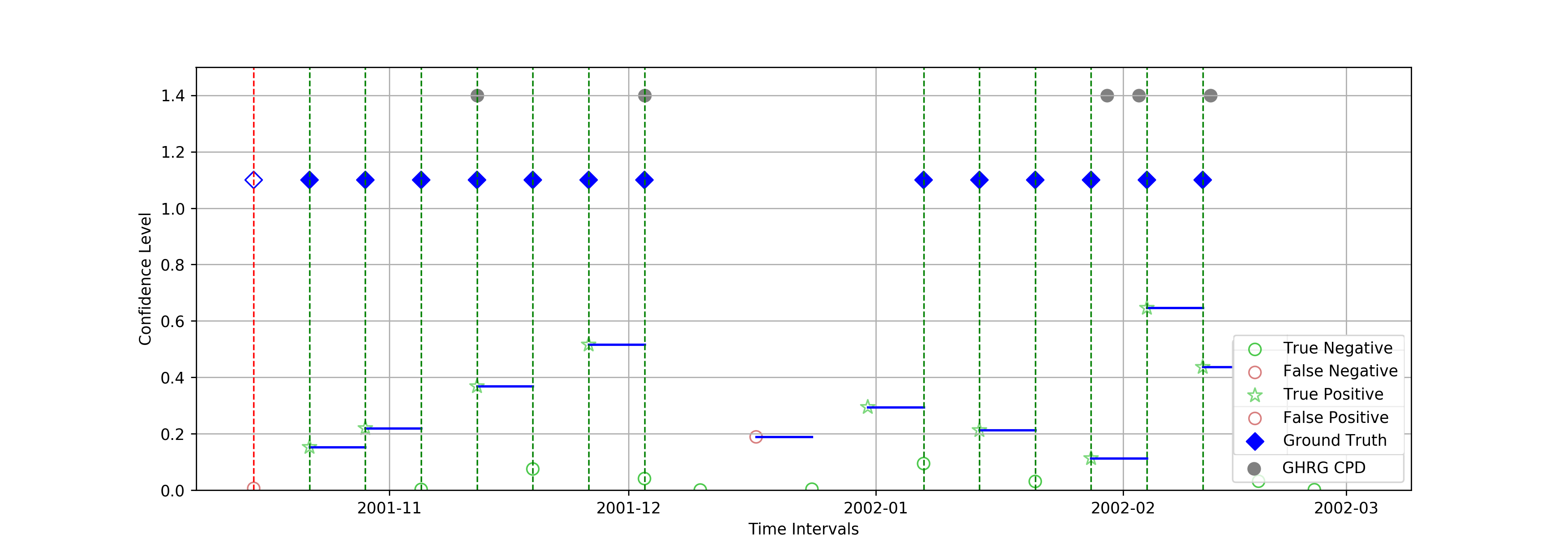}
      \vspace{-7mm}
    \captionsetup{justification=centering,margin=.1cm}
\caption{\small{Enron emails exchange during the second half of 2001, where many events took place. Real events denoted by blue rhombuses, True positive detections by a green star followed by the window length. In grey at the top is the results of the baseline GHRG model~\cite{Peel2015}. Our framework outperforms with recall = 0.9 and Precision = 0.9.}}
   \label{fig:enron}
  \end{minipage}
  \\[12pt]
\begin{minipage}{\textwidth}
  \includegraphics[width=1.1\linewidth]{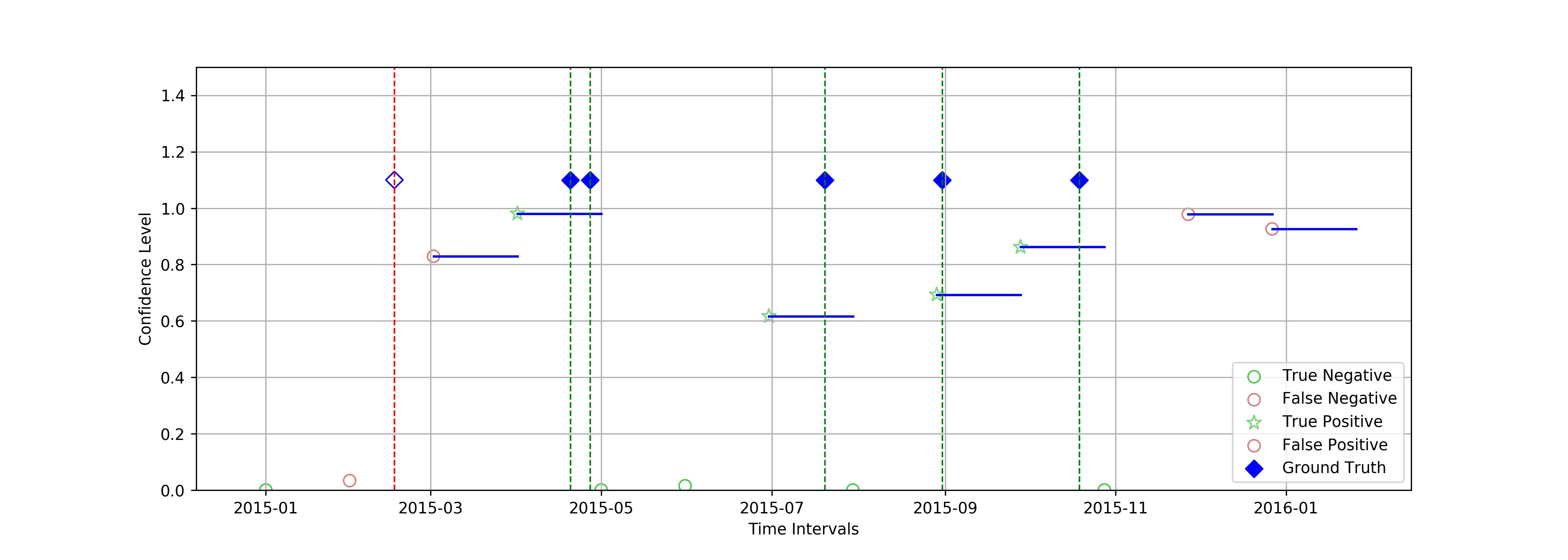}
      \vspace{-7mm}
      \captionsetup{justification=centering,margin=.4cm}
\caption{\small{Ask Ubuntu forum exchange. Release events denoted by blue rhombuses, True positive detections by a green star followed by the window length. Our framework outperforms with recall = 0.8 and Precision = 0.57.}}
  \label{fig:ubuntu}
\end{minipage}
\label{fig:realnets}
\vspace{-.4cm}
\end{figure}
We tested our framework against two real world datasets. The first, the Enron email exchange between 151 employees, mostly managers~\cite{Klimt2004,enronGT}. We generated weekly networks from the emails interactions similar to ~\cite{Peel2015, wang2017fast}. %The ground truth of events is taken from the Enron Timeline~\cite{enronGT}. 
Figure\ref{fig:enron} describes our framework's performance, compared to both the real events, and to the GHRG-based detection framework by Peel and Clauset~\cite{Peel2015}. %Furthermore, we compared our framework to [26] as it is considered one of the states of the art change point detector.  The results are shown in Figure 4. 
Our framework detected 13 out of 14 change points, resulting in during the period of the second half of 2001 where many events effected Enron.  

The second dataset is the interactions on the stack exchange web site Ask Ubuntu~\cite{paranjape2017motifs}, and generated monthly networks. We assume that a new Ubuntu release might affect the community, and extracted the ground truth from the Ubuntu site's detailing release dates.   Figure~\ref{fig:ubuntu}  shows our results. Our framework detected almost all the events with  high confidence: Recall = 0.8, and Precision = 0.57. Clearly, there might be external events that are not version releases that we are not aware of.

\section{Conclusions}
\label{sec:con}
\vspace{-.4cm}
Our framework for size-agnostic detection of changes proved to work across different generative models and real datasets. During the work we have identified am interesting trade-off between precision and recall of detection when considering the size of the network and detectability. We intend to further study this tradeoff in future research. We further plan to try and quantify the nature of the change in the distribution in response to different events.
%\vspace{-.4cm}
%\omitit{
%\vspace{-.4cm}
\begin{acknowledgement}
%\verb|
This research is supported by the Israeli Science Foundation grant \#328/17
%| 
\end{acknowledgement}
%}
\bibliographystyle{spmpsci}
%\bibliography{/Users/ossi/Dropbox/Research/AllCites}

\end{document}